\documentstyle[12pt]{article}
\def\be{\begin{eqnarray}}
\def\ee{\end{eqnarray}}

\def\abstract#1{\vskip 7mm 
\begin{center}{\large Abstract}\par \bigskip
\begin{minipage}[c]{12cm}
\small #1
\end{minipage}
\end{center}
}
\def\title#1{\begin{center}{\Large\bf #1}\end{center}}
\def\author#1{\vskip 5mm \begin{center}{#1}\end{center}}
\def\address#1{\begin{center}{\it #1}\end{center}}
\newcommand{\bfr}{\begin{flushright}}
\newcommand{\efr}{\end{flushright}}
\begin{document}
\vspace*{-0cm}
\title{An Infinite Number of Static Soliton Solutions to 5D Einstein-Maxwell 
Equations}
\author{Takahiro AZUMA\footnote{E-mail: azuma@dokkyo.ac.jp} 
and Takao KOIKAWA\footnote{E-mail: koikawa@otsuma.ac.jp}
}
\vspace{1cm}
\address{
${}^1$ Department of Languages and Culture, Dokkyo University, \\ 
Soka 340-0042, Japan \\
${}^2$ School of Social Information Studies, Otsuma Women's University, \\
Tama 206-8540, Japan
}
\vspace{3.5cm}
\abstract{
The soliton technique is applied to the 5D static Einstein-Maxwell equations, and an infinite number of solutions are explicitly obtained. We study the rod structure of 2-soliton solutions and we show that the 5D Reissner-Nordstr\"om solution and the 5D Majumdar-Papapetrou solution are included as the 2-soliton solutions. 
}

\newpage
\setcounter{page}{2}

\noindent
\section{Introduction}
Ever since the discovery of black ring solution by Emparan and Reall\cite{ER}, 5D black ring solution has attracted considerable attention. It is known that the uniqueness theorem, that there is only one black hole solution for given mass, electric charge and angular momentum, holds in 4D gravity. However the discovery of the black ring solution shows that the theorem does not hold in 5D theory. Therefore, one of the interesting questions is to ask if there can be other black hole solutions in five and higher dimensions. 

In constructing black hole solutions both in four dimensions and in higher dimensions, the soliton technique is an only systematical method. When we write down the vacuum Einstein equation with axial symmetry in terms of the canonical coordinates $(\rho,z)$ in any dimensions\cite{Azenko,Hrmrk}, we find that one of the equations is the zero curvature equation on $\rho$-$z$ plane just like the soliton equations. The 2D soliton equations like KdV equation are to be solved by the inverse scattering method, which shows that these soliton equations are formulated as the zero-curvature equation with a continuity equation. This is the reason why we can also apply the soliton technique to the Einstein equation with axial symmetry. This is also the reason why we can derive an infinite number of soliton solutions in the same way as the soliton equations. In five dimensions, we have already shown that this is really the case for the static case\cite{5DE} and stationary case\cite{5DRE}. Among the infinite number of solutions, both regular and singular solutions are included. In order to extract physically interesting solutions, the study of the rod structure\cite{Hrmrk} is powerful. The fact that the soliton solutions given by the inverse scattering method are constructed by ``poles" $\mu_k$ fits to studying the rod structure of the solutions.

The black ring solution can exist on the balance of the gravitational force and the centrifugal force. Then it is natural to expect a solution in which the gravitational force balances the electric force in five dimensions. Several works\cite{CRBR}-\cite{5DEMD} devoted to the study of black ring solutions with electric charge. In this paper, we study the 5D static Einstein-Maxwell(EM) equations with axial symmetry. The equations at first sight do not seem to reduce to the zero-curvature equation. However, after introducing new functions, we can show that they are also cast into a zero-curvature equation and some other associated equations, which enables one to find an infinite number of solutions also in the presence of electric field.

This paper is constructed as follows. In the next section we derive an infinite number of soliton solutions for the 5D static EM equations. In section 3, we focus on 2-soliton solutions and study the rod structure of the solutions. By evading the conical singularity from the solution, we obtain the 5D Reissner-Nordstr\"om(RN) solution and the 5D Majumdar-Papapetrou(MP) solution\cite{MP}. The last section is devoted to summary and discussion. 

%
%
\section{Solutions of Einstein-Maxwell equations}
The metric we consider is given by
\be
ds^2=f(d\rho^2+dz^2)+g_{ab}dx^adx^b, \quad (a,b=0,1,2)
\ee
where $f$ and $g_{ab}$ are functions of $\rho$ and $z$. The 5D EM equations read
\be
&&R^\mu{}_\nu=2\left( F^{\mu\alpha}F_{\nu\alpha}
-\frac{1}{6}\delta^\mu{}_\nu
F^{\alpha\beta}F_{\alpha\beta} \right),\\
&&F^{\mu\nu}{}_{;\mu}=0,
\ee
with
\be
&&F_{\mu\nu}=A_\nu,_\mu-A_\mu,_\nu.
\ee
We solve these equations under the static condition with the coordinate condition $\det g=-\rho^2$. We also assume that there is only electric charge. Then, the part of the metric $g=(g_{ab})$ and the $U(1)$ gauge field are assumed to have the fallowing form:
\be
&&g={\rm diag}\left(-h_1^{-1}h_2^{-1},
\left[\sqrt{\rho^2+z^2}-z\right]h_1,
\left[\sqrt{\rho^2+z^2}+z\right]h_2\right),
\\
&&A_0=-\chi,\quad A_1=A_2=A_\rho=A_z=0,
\ee
where $h_1$ and $h_2$ are functions of $\rho$ and $z$, and $\chi(\rho,z)$ is the electrostatic potential.
Then the EM equations are explicitly written as
\be
&&[\rho (\ln h_i),_\rho],_\rho+[\rho (\ln h_i),_z],_z
=-\frac{4\rho}{3}h_1h_2(\chi,_\rho^2+\chi,_z^2), \quad (i=1,2)\label{em1}\\
\medskip
&&(\ln f),_\rho+\frac{\rho}{\rho^2+z^2}-\frac{1}{2}\left(
\frac{\sqrt{\rho^2+z^2}+z}{\sqrt{\rho^2+z^2}}(\ln h_1),_\rho+
\frac{\sqrt{\rho^2+z^2}-z}{\sqrt{\rho^2+z^2}}(\ln h_2),_\rho \right.\cr
&&+
\rho[(\ln h_1),_\rho^2+(\ln h_2),_\rho^2+(\ln h_1),_\rho(\ln h_2),_\rho]+
\frac{\rho}{\sqrt{\rho^2+z^2}}[(\ln h_1),_z-(\ln h_2),_z] \cr
&&-
\rho[(\ln h_1),_z^2+(\ln h_2),_z^2+(\ln h_1),_z(\ln h_2),_z]
\Biggr)=-2\rho h_1h_2(\chi,_\rho^2-\chi,_z^2),\label{em2}\\
\medskip
&&(\ln f),_z+\frac{z}{\rho^2+z^2}-\frac{1}{2}\left(
\frac{\sqrt{\rho^2+z^2}+z}{\sqrt{\rho^2+z^2}}(\ln h_1),_z+
\frac{\sqrt{\rho^2+z^2}-z}{\sqrt{\rho^2+z^2}}(\ln h_2),_z \right.\cr
&&-
\frac{\rho}{\sqrt{\rho^2+z^2}}[(\ln h_1),_\rho-(\ln h_2),_\rho]+
2\rho[(\ln h_1),_\rho (\ln h_1),_z+(\ln h_2),_\rho (\ln h_2),_z] \cr
&&+
\rho[(\ln h_1),_\rho (\ln h_2),_z+(\ln h_1),_z (\ln h_2),_\rho]
\Biggr)=-4\rho h_1h_2\chi,_\rho\chi,_z,\label{em3}\\
\medskip
&&(\rho h_1h_2\chi,_\rho),_\rho+(\rho h_1h_2\chi,_z),_z=0.\label{em4}
\ee
Here we assume the following ansatz to solve these equations:
\be
h_1&=&\left( 1-\frac{8}{3}c\chi+\frac{4}{3}\chi^2\right)^{-1/2}N^{1/2},\label{h1}\\
h_2&=&\left( 1-\frac{8}{3}c\chi+\frac{4}{3}\chi^2\right)^{-1/2}N^{-1/2},\label{h2}
\ee
where $N$ is function of $\rho$ and $z$, and $c$ is a constant. Then the equations (\ref{em1}) and (\ref{em4}) are both put into
\be
\chi,_{\rho\rho}+\chi,_{zz}+\rho^{-1}\chi,_\rho
=8(3-8c\chi+4\chi^2)^{-1}(\chi-c)(\chi,_\rho^2+\chi,_z^2),\label{chi}
\ee
with
\be
(\ln N),_{\rho\rho}+(\ln N),_{zz}+\rho^{-1}(\ln N),_\rho=0.\label{lap1}
\ee
The last equation is the Laplace equation for $\ln N$, and so we can solve this equation. The equation (\ref{em2}) is written as
\be
&&(\ln f),_\rho=-\frac{\rho}{\rho^2+z^2}-\frac{4(\chi-c)\chi,_\rho}
{3-8c\chi+4\chi^2}+\frac{6\rho(4c^2-3)(\chi,_\rho^2-\chi,_z^2)}
{(3-8c\chi+4\chi^2)^2}\cr
&&+\frac{1}{2\sqrt{\rho^2+z^2}}\left[z(\ln N),_\rho+\rho(\ln N),_z\right]
+\frac{\rho}{8}\left[(\ln N),_\rho^2-(\ln N),_z^2\right],\label{fx}
\ee
and the equation (\ref{em3}) as
\be
&&(\ln f),_z=-\frac{z}{\rho^2+z^2}-\frac{4(\chi-c)\chi,_z}
{3-8c\chi+4\chi^2}+\frac{12\rho(4c^2-3)\chi,_\rho\chi,_z}
{(3-8c\chi+4\chi^2)^2}\cr
&&-\frac{1}{2\sqrt{\rho^2+z^2}}\left[\rho(\ln N),_\rho-z(\ln N),_z\right]
+\frac{\rho}{4}(\ln N),_\rho(\ln N),_z.\label{fz}
\ee
We further introduce a new function $R(\rho, z)$ by
\be
\chi=\frac{e}{R+m},
\ee 
where $e$ and $m$ are constants satisfying
\be
m=\frac{4}{3}ce.
\ee
Then the equations (\ref{h1}) and (\ref{h2}) lead to
\be
h_1&=&\frac{m+R}{\sqrt{R^2-d^2}}N^{1/2},\label{h1a}\\
h_2&=&\frac{m+R}{\sqrt{R^2-d^2}}N^{-1/2},\label{h2a}
\ee
where
\be
d=\sqrt{\frac{3m^2-4e^2}{3}}.
\ee
The equation (\ref{chi}) is also expressed in terms of $R$ as
\be
R,_{\rho\rho}+R,_{zz}+\rho^{-1}R,_\rho=2R(R^2-d^2)^{-1}(R,_\rho^2+R,_z^2).
\ee
Introducing a function $h(\rho, z)$ by
\be
R=d\frac{1+h}{1-h},
\ee
we can rewrite this equation as
\be
(\ln h),_{\rho\rho}+(\ln h),_{zz}+\rho^{-1}(\ln h),_\rho=0, \label{lap2}
\ee
which is the Laplace equation for $\ln h$. By further introducing $Q(\rho, z)$ by
\be
f=\frac{1}{2\sqrt{\rho^2+z^2}}
\left( 1-\frac{8}{3}c\chi+\frac{4}{3}\chi^2\right)^{-1/2}Q,
\ee
the equations (\ref{fx}) and (\ref{fz}) are to be expressed as
\be
&&(\ln Q),_\rho=\frac{3\rho}{8}[(\ln h),_\rho^2-(\ln h),_z^2]\cr
&&+\frac{1}{2\sqrt{\rho^2+z^2}}\left[z(\ln N),_\rho+\rho(\ln N),_z\right]
+\frac{\rho}{8}\left[(\ln N),_\rho^2-(\ln N),_z^2\right],\label{qx}
\ee
and
\be
&&(\ln Q),_z=\frac{3\rho}{4}(\ln h),_\rho(\ln h),_z\cr
&&-\frac{1}{2\sqrt{\rho^2+z^2}}\left[\rho(\ln N),_\rho-z(\ln N),_z\right]
+\frac{\rho}{4}(\ln N),_\rho(\ln N),_z.\label{qz}
\ee
Now that we have two Laplace equations in (\ref{lap1}) and (\ref{lap2}) for $\ln N$ and $\ln h$ respectively, we assume that
\be
N=h.
\ee
Then equations (\ref{qx}) and (\ref{qz}) are reduced to
\be
(\ln Q),_\rho=\frac{\rho}{2}[(\ln h),_\rho^2-(\ln h),_z^2]
+\frac{1}{2\sqrt{\rho^2+z^2}}\left[z(\ln h),_\rho+\rho(\ln h),_z\right],
\ee
and
\be
(\ln Q),_z=\rho(\ln h),_\rho(\ln h),_z
-\frac{1}{2\sqrt{\rho^2+z^2}}\left[\rho(\ln h),_\rho-z(\ln h),_z\right].
\ee
This shows that we can integrate these equations to obtain $Q$ by using the solution $h$ that is obtained by solving the Laplace equations.

There can be solitonic and non-soliotnic solution to the Laplace equation. Here we adopt the solitonic solutions for $h$ and $Q$. The $n$-soliton solution is given by
\be
h&=&\prod_k^{n}(i\mu_k/\rho)^\delta,\\
Q&=&\left[\frac{\rho^{n^2/2}\prod_{k>l}^n(\mu_k-\mu_l)^2}
{\prod_k^n(\rho^2+\mu_k^2)\prod_l^n\mu_l^{n-2}C^{(n)}}\right]^{\delta^2}\cr
&\times&\left[\left(\prod_k^ni\frac{\mu_k}{\rho}\right)^{1/2}
\frac{\left(z+\sqrt{\rho^2+z^2}\right)^{n/2}\prod_k^{n/2}(2w_{2k})}
{\prod_k^n\left(\mu_k+z+\sqrt{\rho^2+z^2}\right)}\right]^\delta,
\ee
with
\be
&&\mu_k=w_k-z+(-1)^{k-1}\sqrt{(w_k-z)^2+\rho^2},\\
&&C^{(n)}=2^{n(n-2)/2}\prod_{k>l}^{n/2}
(w_{2k-1}-w_{2l-1})^2(w_{2k}-w_{2l})^2,
\ee
where $\delta$ and $w_k(k=1,2,\cdots,n)$ are constants.

In order to elucidate the solutions, we illustrate the 2-soliton solution by setting $\delta=1$. In this case $h$ and $Q$ are given by
\be
h&=&-\frac{\mu_1\mu_2}{\rho^2},\\
Q&=&\frac{2(z_0+d)\rho\left(z+\sqrt{\rho^2+z^2}\right)\sqrt{-\mu_1\mu_2}
(\mu_1-\mu_2)^2}
{\left(\mu_1+z+\sqrt{\rho^2+z^2}\right)\left(\mu_2+z+\sqrt{\rho^2+z^2}\right)
(\mu_1^2+\rho^2)(\mu_2^2+\rho^2)},
\ee
where we have put $w_1=z_0-d$ and $w_2=z_0+d$, and so $\mu_1$ and $\mu_2$ are given by
\be
\left\{
\matrix{
\mu_1=z_0-z-d+\sqrt{(z_0-z-d)^2+\rho^2},\cr
\mu_2=z_0-z+d-\sqrt{(z_0-z+d)^2+\rho^2}.}
\right.
\ee
We rewrite these poles by transforming $z$ to $z+z_0$ as
\be
\left\{
\matrix{
\mu_1=-(z+d)+\sqrt{(z+d)^2+\rho^2},\cr
\mu_2=-(z-d)-\sqrt{(z-d)^2+\rho^2}.}
\right.
\ee
By using these expressions, the 2-soliton solution is given by
\be
&&g_{00}=\frac{4d^2\mu_1\mu_2\rho^2}
{[(d+m)\rho^2-(d-m)\mu_1\mu_2]^2},\\
&&g_{11}=\mu^*h_1,\\
&&g_{22}=-\mu h_2,\\
&&f=-\frac{(z_0+d)\mu[(d+m)\rho^2-(d-m)\mu_1\mu_2](\mu_1-\mu_2)^2}
{2d\sqrt{(z+z_0)^2+\rho^2}(\mu_1-\mu)(\mu_2-\mu)(\mu_1^2+\rho^2)(\mu_2^2+\rho^2)},
\ee
where $\mu$ and $\mu^*$ are
\be
\left\{
\matrix{
\mu=-(z+z_0)-\sqrt{(z+z_0)^2+\rho^2},\cr
\mu^*=-(z+z_0)+\sqrt{(z+z_0)^2+\rho^2},}
\right.
\ee
and $h_1$ and $h_2$ in (\ref{h1a}) and (\ref{h2a}) are here given by 
\be
h_1&=&\frac{(d+m)\rho^2-(d-m)\mu_1\mu_2}{2d\rho^2},\cr
h_2&=&-\frac{(d+m)\rho^2-(d-m)\mu_1\mu_2}{2d\mu_1\mu_2}.
\ee

%
%
\section{Analysis of Solutions}
In analyzing the behavior of the metric, we use a notion of rod structure. We recapitulate it here following the discussion by Harmark\cite{Hrmrk}. A rod is defined as an interval $[a_{k-1}, a_{k}]~(k=1,2,\dots,N+1)$ on the $z$-axis where the parameters $a_k$ are assumed to be in the order
\be
a_0=-\infty<a_1<...<a_N<a_{N+1}=\infty.
\ee
The rod structure is specified by these intervals and $N+1$ non-zero vectors $v_{(k)}~(k=1,2,\dots,N+1)$ in the intervals. In the intervals $[a_{k-1}, a_{k}]~(k=1,2,\dots,N+1)$, the vectors $v_{(k)}$ are defined as the eigen-vectors belonging to zero eigen-value by
\be
g v_{(k)}=0. \quad (k=1,2,\dots,N+1)
\ee
Here we should note that, because of the auxiliary condition $\det g=-\rho^2$,  at least one of the eigen-values is zero on the $z$-axis. The necessary condition for the space-time to be free from a singularity, there should be only one eigen-vector belonging to zero eigen-value in each interval. 
Each rod is called space-like or time-like according as $g_{ab}v^av^b/\rho^2>0$ or $g_{ab}v^av^b/\rho^2<0$, respectively. When the rod is time-like, this rod corresponds to a horizon or curvature singularity. When the rod is space-like and singularity-free, we need to study whether there is a conical singularity or not.

We confine ourselves to the diagonal $g$ case, since we here discuss the static case. A vector in three dimensions is spanned by three bases, which we denote by $\partial/\partial x^a~(a=0,1,2)$. Suppose that $g_{aa}$ of the diagonal $g$ components approaches zero as $\rho \to 0$, then the non-zero eigen-vector belonging to zero eigen-value is in the $\partial/\partial x^a$ direction, which is called the direction of the rod. When the rod is space-like, the coordinates $x^i~(i=1,2)$ is a compact spatial direction and so there should be a period. When we go around the $z$-axis on $\rho$-$i$ plane perpendicular to the $z$-axis, the period $\triangle x^i$ should be
\be
\triangle x^i=2\pi \lim_{\rho \to 0}\sqrt{\frac{\rho^2 f}{g_{ii}v^iv^i}}=2\pi \lim_{\rho \to 0}\sqrt{\frac{\rho^2 f}{g_{ii}}},
\ee
in order to be free from a conical singularity. This will be used as a criterion in finding regular solutions. 

In order to study the asymptotic behaviors of the metric components, we expand them in the asymptotic region where $\sqrt{\rho^2+z^2}\to\infty$
($h_1\to 1,\ h_2\to 1$). We have
\be
&&g_{00}\sim -1,\quad g_{11}\sim -z+\sqrt{\rho^2+z^2},\cr
&&g_{22}\sim z+\sqrt{\rho^2+z^2},\quad
f\sim \frac{1}{2\sqrt{\rho^2+z^2}},
\ee
which show the asymptotic flatness. 

We next study $\rho\sim 0$ behaviors of the metric components, or the rod structure of the solutions.  As $z=z_0$ does not give any difference in the rod structures, we divide the $z$-axis into four intervals separated by values $z=-z_0$ and $z=\pm d$. We assume that these parameters are in the following order without loss of generality:
\be
-\infty<-d<d<-z_0<\infty.
\ee
When $g_{ii} \propto \rho^2~(i=1,2)$ in some interval, the rod is space-like and in $\partial/\partial x^i$ direction and $\rho$-$i$ plane is perpendicular to the $z$-axis. Going around the $z$-axis in the $i$-th direction, we study the period of $x^i$ coordinate. The behaviors of the metric components at $\rho\sim 0$ in each region are given as follows.\\
\noindent
(i) $\infty>z>-z_0$
\be
g_{00}&\sim&-\frac{z^2-d^2}{(z+m)^2},\\
g_{11}&\sim&\frac{\rho^2(z+m)}{2(z+d)(z+z_0)},\\
g_{22}&\sim&\frac{2(z+z_0)(z+m)}{z-d},\\
f&\sim&\frac{z+m}{2(z+d)(z+z_0)}.
\ee
In this interval, we note that $g_{11}\sim {\cal O}(\rho^2)$. Since the rod is in $\partial/\partial x^1$ direction, we study the period $\triangle x^1$ to find that
\be
\triangle x^1=2\pi \lim_{\rho \to 0}\sqrt{\frac{\rho^2f}{g_{11}}}&=2\pi,
\ee
which shows no conical singularity in this interval.\\
\noindent
(ii) $-z_0>z>d$
\be
g_{00}&\sim&-\frac{z^2-d^2}{(z+m)^2},\\
g_{11}&\sim&-\frac{2(z+m)(z+z_0)}{z+d},\\
g_{22}&\sim&-\frac{\rho^2(z+m)}{2(z-d)(z+z_0)},\\
f&\sim&-\frac{(z+m)(z_0+d)}{2(z-d)(z+z_0)(z_0-d)}.
\ee
In this interval, we note that $g_{22}\sim {\cal O}(\rho^2)$, and the rod is in $\partial/\partial x^2$ direction. So we study the period $\triangle x^2$ to find that
\be
\triangle x^2=2\pi \lim_{\rho \to 0}\sqrt{\frac{\rho^2f}{g_{22}}}&=2\pi\sqrt{\displaystyle{\frac{z_0+d}{z_0-d}}},
\ee
which shows that two parameters $z_0$ and $d$ should be adjusted to avoid the conical singularity.\\
\noindent
(iii) $d>z>-d$
\be
g_{00}&\sim&\frac{d^2\rho^2}{(z^2-d^2)(d+m)^2},\\
g_{11}&\sim&-\frac{(z+z_0)(d+m)}{d},\\
g_{22}&\sim&\frac{(z^2-d^2)(d+m)}{d(z+z_0)},\\
f&\sim&\frac{d(d+m)}{(z^2-d^2)(z_0-d)}.
\ee
As $g_{00}\sim {\cal O}(\rho^2)$ in this interval, the eigen-vector belonging to the zero eigen-value is given by $v=(1,0,0)$. Then we obtain that $g_{ab}v^av^b/\rho^2=g_{00}/\rho^2=d^2/((z^2-d^2)(d+m)^2)<0$. This shows that the rod is time-like, and so this interval is the horizon.\\
\noindent
(iv) $-d>z>-\infty$
\be
g_{00}&\sim&-\frac{z^2-d^2}{(z-m)^2},\\
g_{11}&\sim&-\frac{2(z-m)(z+z_0)}{z-d},\\
g_{22}&\sim&-\frac{\rho^2(z-m)}{2(z+d)(z+z_0)},\\
f&\sim&-\frac{z-m}{2(z+d)(z+z_0)}.\\
\ee
In this interval, we note that $g_{22}\sim {\cal O}(\rho^2)$ as in the case (ii), and the period $\triangle x^2$ is given by
\be
\triangle x^2=2\pi \lim_{\rho \to 0}\sqrt{\frac{\rho^2f}{g_{22}}}&=2\pi,
\ee
which shows that there is no conical singularity.

In Table 1, the rod structure is summarized. The rod in $[-d,~d]$ representing the event horizon is sandwiched by the rods in the same $\frac{\partial}{\partial x^2}$ direction, which shows that the topology of the event horizon is $S^2 \times S^1$ as in the static black ring case.
\begin{table}
\begin{center}
\begin{tabular}{|c|c|c|c|c|}\hline
Rods &$ [-\infty,-d] $&$  [-d,d]$&$ [d,-z_0]$&$ [-z_0,\infty]$ \\ \hline

Directions &$ \frac{\partial}{\partial x^2}$&$ \frac{\partial}{\partial x^0}$&$ \frac{\partial}{\partial x^2}$&$ \frac{\partial}{\partial x^1}$ \\ \hline

Periods &$ 2 \pi $&$ \rm{Event~Horizon}$&$ 2 \pi \sqrt{\frac{z_0+d}{z_0-d}}$&$ 2 \pi$ \\ \hline

\end{tabular}
\caption{Rod structure of the 2-soliton solution to the 5D static EM equation}
\end{center}
\begin{center}
\begin{tabular}{|c|c|c|c|c|}
\hline
Rods &$ [-\infty,-\sigma] $&$  [-\sigma,\sigma]$&$ [\sigma,-z_0]$&$ [-z_0,\infty]$ \\ \hline
Directions &$ \frac{\partial}{\partial x^2}$&$ \frac{\partial}{\partial x^0}$&$ \frac{\partial}{\partial x^2}$&$ \frac{\partial}{\partial x^1}$ \\ \hline
Periods &$ 2 \pi $&$ \rm{Event~Horizon}$&$ 2 \pi \sqrt{\frac{z_0+\sigma}{z_0-\sigma}}$&$ 2 \pi$ \\ \hline
\end{tabular}
\caption{Rod structure of the static black ring solution}
\end{center}
\end{table}

We find that there is a conical singularity in $[d, -z_0]$. In order to get rid of the conical singularity, there can be two ways.  The first is to set $d=-z_0$, which gets rid of the interval itself. The second is to make $\triangle x^2= 2\pi$ by imposing $d=0$. This should be compared with the case of static black ring, where there is no electric charge. The rod structure is depicted in Table 2\cite{5DE}.This shows that the conical singularity in $[\sigma, -z_0]$ has the same structure as in the present case. When we set $\sigma=0$, the metric becomes flat. However, we should note that, in the present case, the metric is not reduced to the flat one as $d \to 0$. Therefore, we need to investigate it. The $d=0$ case is studied by further imposing $z_0\ne 0$ or $z_0=0$. 

We first compute the $d=-z_0(\mu_2=\mu)$ case. In this case, the metric components are given by
\be
g_{00}&=&\frac{4d^2\mu_1\mu\rho^2}
{[(d+m)\rho^2-(d-m)\mu_1\mu]^2},\\
g_{11}&=&\mu^*h_1,\\
g_{22}&=&-\mu h_2,\\
f&=&\frac{[(d+m)\rho^2-(d-m)\mu_1\mu](\mu_1-\mu)}
{2d(\mu_1^2+\rho^2)(\mu^2+\rho^2)},
\ee
where 
\be
h_1&=&\frac{(d+m)\rho^2-(d-m)\mu_1\mu}{2d\rho^2},\\
h_2&=&-\frac{(d+m)\rho^2-(d-m)\mu_1\mu}{2d\mu_1\mu}.
\ee
We shall rewrite these metric components by introducing $(\rho,\theta)$ coordinates by 
\be
\rho&=&\frac{1}{2}\sqrt{(r^2-2m)^2-4d^2}\sin 2\theta,\\
z&=&\frac{1}{2}(r^2-2m)\cos 2\theta.
\ee
Noting that $h_1$ and $h_2$ are rewritten as
\be
h_1&=&\frac{r^2}{r^2-2m+2d},\\
h_2&=&\frac{r^2}{r^2-2m-2d},
\ee
the metric components are reduced to
\be
g_{00}&=&-\frac{(r^2-2m)^2-4d^2}{r^4},\\
g_{11}&=&r^2\sin^2\theta,\\
g_{22}&=&r^2\cos^2\theta,\\
f&=&\frac{r^2}{(r^2-2m)^2-4d^2\cos^22\theta}.
\ee
Using the relation
\be
d\rho^2+dz^2=\left[(r^2-2m)^2-4d^2\cos^22\theta\right]
\left[\frac{r^2dr^2}{(r^2-2m)^2-4d^2}+d\theta^2\right],
\ee
we find that the metric is the 5D RN solution given by
\be
ds^2=&-&\left(1-\frac{4m}{r^2}+\frac{16e^2}{3r^4}\right)dt^2
+\left(1-\frac{4m}{r^2}+\frac{16e^2}{3r^4}\right)^{-1}dr^2 \cr
\medskip
&+&r^2d\theta^2+r^2\sin^2\theta d\phi^2+r^2\cos^2\theta d\psi^2.
\ee

Next we study the $d=0$ case. This is the extremal case satisfying $m=2/\sqrt{3}e$ and the solution is MP type. When $z_0 \ne 0$ we have the metric components
\be
&&g_{00}=-\frac{\rho^2+z^2}{(\sqrt{\rho^2+z^2}+m)^2},\\
&&g_{11}=\frac{\left[\sqrt{\rho^2+(z+z_0)^2}-(z+z_0)\right]
(\sqrt{\rho^2+z^2}+m)}
{\sqrt{\rho^2+z^2}},\\
&&g_{22}=\frac{\left[\sqrt{\rho^2+(z+z_0)^2}+(z+z_0)\right]
(\sqrt{\rho^2+z^2}+m)}
{\sqrt{\rho^2+z^2}},\\
&&f=\frac{\sqrt{\rho^2+z^2}+m}
{2\sqrt{\rho^2+(z+z_0)^2}\sqrt{\rho^2+z^2}}.
\ee
Although the parameter $d$, or the radius of $S^2$, becomes zero, the solution has the remaining ring-like structure at $z=0$. We shall show that this ring becomes the curvature singularity. In this limit, both $g_{00}$ and $g_{22}$ are proportional to $\rho^2$ at $z=0$. This implies that there are two independent eigen-vectors belonging to zero eigen-value, and so there should appear a singularity. Actually, by using these metric components, we find the curvature invariant is evaluated at $z=0$ as
\be
R^{\alpha\beta\gamma\delta}R_{\alpha\beta\gamma\delta}\sim 
\frac{31z_0^2}{\rho^2m^2}.
\ee

Last of all, we study the $d=z_0=0$ case. In this case, we have
\be
&&g_{00}=-\frac{\rho^2+z^2}{(\sqrt{\rho^2+z^2}+m)^2},\\
&&g_{11}=\frac{\left(\sqrt{\rho^2+z^2}-z\right)
(\sqrt{\rho^2+z^2}+m)}{\sqrt{\rho^2+z^2}},\\
&&g_{22}=\frac{\left(\sqrt{\rho^2+z^2}+z\right)
(\sqrt{\rho^2+z^2}+m)}{\sqrt{\rho^2+z^2}},\\
&&f=\frac{\sqrt{\rho^2+z^2}+m}
{2(\rho^2+z^2)}.
\ee
We introduce the $(\rho,\theta)$ coordinates by
\be
\rho&=&\frac{1}{2}(r^2-2m)\sin 2\theta, \\
z&=&\frac{1}{2}(r^2-2m)\cos 2\theta.
\ee
Then we find that the above metric turns out to be that of the 5D MP solution given by
\be
ds^2=&&-\left(1-\frac{2m}{r^2}\right)^2dt^2
+\left(1-\frac{2m}{r^2}\right)^{-2}dr^2+r^2d\theta^2 \cr
\medskip
&&+r^2\sin^2\theta d\phi^2+r^2\cos^2\theta d\psi^2.
\ee
%
%
\section{Summary and Discussion}
In this paper we obtain an infinite number of static solutions to the 5D EM equations. The solutions are examined in detail for the 2-soliton solution case. By studying the rod structures of the solutions, we get rid of the conical singularity from the general 2-soliton solution. As a result, we obtain the 5D RN and MP solution as 2-soliton solutions. 

The 5D rotating black ring solution is the solution where the gravity force and the centrifugal force are in equilibrium. When an electric charge is taken into account, it is expected that the repulsive electric force might balance the attractive gravity force. As far as we study the 2-soliton solutions, the black ring topology $S^2 \times S^1$ without conical singularity does not appear. This might be found in higher number of soliton solutions, or it might be found to by constructing solitons from a non-flat background as in \cite{TN}. This would be reported in future.

\end{document}